\documentstyle[sprocl]{article}

\def\Journal#1#2#3#4{{#1} {\bf #2}, #3 (#4)}

\def\NCA{\em Nuovo Cimento}

\def\NPB{{\em Nucl. Phys.} B}
\def\PLB{{\em Phys. Lett.}  B}

\def\PR{\em Phys. Rev.}
\def\PRD{{\em Phys. Rev.} D}

\def\be{\begin{equation}}
\def\ee{\end{equation}}
\def\ba{\begin{array}}
\def\ea{\end{array}}
\def\bea{\begin{eqnarray}}
\def\eea{\end{eqnarray}}
\def\slask{\!\!\!\!/}
\def\g{\gamma}
\def\eps{\epsilon}
\def\cA{{\cal A}}

\def\tr{{\rm tr\,}}
\def\nn{\nonumber\\}
\def\goto{\rightarrow}

\def\bfT{{\bf T }}
\def\at#1#2{\left.#1\right|_{#2}}

\begin{document}

\title{ANOMALIES AT HIGH TEMPERATURE}

\author{P. ELMFORS}

\address{Stockholm University, Fysikum, Box 6730, S-113 85 Stockolm,
	Sweden}


\maketitle\abstracts{The anomaly equation can be derived from the
ultraviolet properties of quantum field theory and should, therefore, not
depend on infrared properties, such as the presence of a thermal heat bath. 
There is also an infrared explanation of 
anomalies which is related to fermionic zero modes. 
I show how the anomaly equation can be satisfied in a high temperature 
plasma in spite of the fact that all propagating fermionic 
excitations have a thermal mass. }
  
\section{Introduction}
\begin{figure}[t]
\setlength{\unitlength}{1mm}
\begin{picture}(50,80)(0,0)
\includegraphics{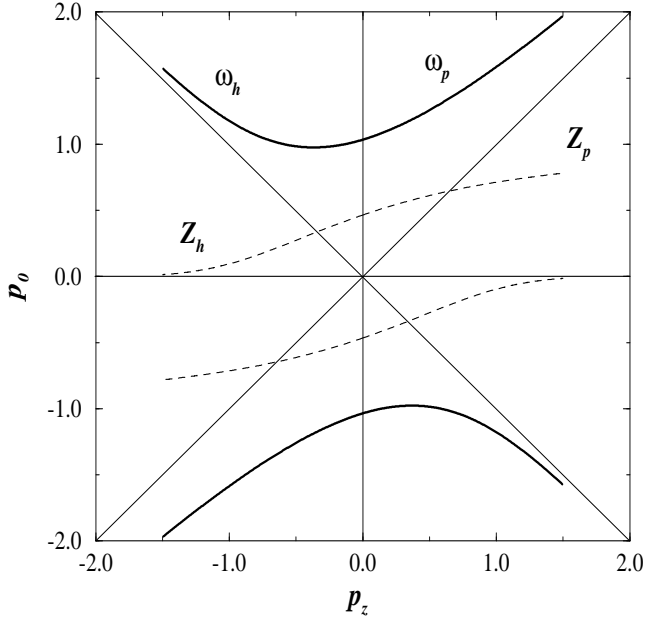}
   \put(0,0){}
\end{picture}
\caption{Dispersion relations $\omega_{p,h}$ and spectral weights $Z_{p,h}$
for a right-handed (anti) particle/hole in the lowest Landau level with
a magnetic field in the positive $z$ direction.
\label{f:n0x2}}
\end{figure}

The anomaly of the conservation of baryon and lepton number currents 
is of obvious importance for baryogenesis. Though it can be derived purely
from the UV behaviour of fermions in a background field it is reassuring to
understand it also 
from a more intuitive IR point of view.\cite{Elmfors97} In vacuum there
is a close relation between the existence of zero modes and the anomaly,
most conveniently formulated in terms of the index of the Dirac operator in
a background field. A physical picture is that particles are pumped up from
the Dirac sea and at the same time holes, representing antiparticles, move
down into the Dirac sea.\cite{NielsenN83} In vacuum 
this can only happen continuously if
energy levels cross the Dirac surface.
At finite temperature all propagating particles have a finite
thermal mass of order $gT$ and there is no level crossing. 
It is therefore at first difficult to imagine how thermal
particles can be created without overcoming the mass gap. We should keep in
mind that an ordinary Dirac mass gap suppresses the creation of particles
exponentially. If this were the case also for the effective 
thermal mass, which is not a local mass of Dirac type, the production of
fermions from sphalarons could have been suppressed. To resolve the 
paradox it is absolutely essential to abandon the strict
quasi-particle picture. The reason why
the anomaly equation is satisfied even in presence
of thermal masses, where the  standard  level crossing picture is  not
applicable  since no  levels ever  cross  the Dirac surface, is that the
spectral weight $Z$ varies continuously from zero to  one on each
particle/hole branch (Fig.~\ref{f:n0x2}). Thus the topological argument of
level crossing does not apply at finite temperature.
I  show explicitly,
using the full spectral function in a background of electric and magnetic
fields, how   the  produced chiral charge only depends
on the ultraviolet properties of the spectral function.
I do this in the context of the
chiral anomaly in massless QED using the Hard Thermal Loop (HTL) effective
action to describe the dynamics of the fermions at high temperature.
\vspace{-1cm}
\section{HTL spectral function in a magnetic field}
The HTL effective action for
QED can be written as \cite{BraatenP90}
\begin{eqnarray}\label{HTLEA}
    {\cal L}_{\rm HTL}&=&-{1 \over 4}F^2+\frac{3}{4}{\cal M}^2_\gamma
    F_{\mu\alpha}\left<\frac{u^\alpha u^\beta}
      {(\partial \cdot u)^2}\right> F_\beta^{~~\mu}\nn
      && +\,\,\overline{\Psi}(\Pi\slask-m)\Psi
    -{\cal M}_e^2\overline{\Psi}\gamma_\mu\left<\frac{u^\mu}{u\cdot\Pi}
      \right>\Psi~~,
\end{eqnarray}
where $\Pi_\mu=i\partial_\mu-gA_\mu$ and
the average $\left<\cdot\right>$ is defined by
\begin{equation}\label{ave}
    \left< f(u_0,\vec{u})\right>=\int \frac{d\Omega}{4\pi}
      f(u_0,\vec{u})~~,
\ee
where $u_0=1$ and  $\vec{u}$ is a   spatial unit vector.  The  thermal
mass of the  photon ${\cal M}^2_\gamma$ is  given by $g^2T^2/9$ and for the
electron  we have  ${\cal M}_e^2=g^2T^2/8$.  The  equation   of motion  for
$\Psi$    that follows      from    Eq.~(\ref{HTLEA})    is
\begin{equation}\label{eqom}
    \left[\Pi\slask-m-{\cal M}_e^2
     \gamma_\mu
     \left<\frac{u^\mu}{u\cdot\Pi}\right>\right]\Psi=0~~.
\ee
Equation
(\ref{eqom})  is a  non-local  and  non-linear differential  equation,
which  is, in   general, very  difficult  to solve.   What makes  this
equation much less tractable than  the  thermal Dirac equation, in  the
absence of an external electromagnetic  field, is that the average over
$\vec{u}$     is        difficult       to          perform explicitly since
$[\Pi_\mu,\Pi_\nu]=-igF_{\mu\nu}\neq  0$,\  i.~e.  not all components of
$\Pi_\mu$   can  be diagonalized  simultaneously.    We  shall first 
only deal with an external magnetic field and fix  it to be in
the $z$-direction. The solutions to Eq.~(\ref{eqom}) in vacuum 
(${\cal M}_e=0$) are
given by the standard Landau levels.  Since  the spatial symmetries of
the system are unchanged by the thermal heat bath, we expect the
eigenfunctions to   have    the   same  spatial   form    as  at  zero
temperature. In fact,  after performing the  $u$-integral in Eq.~(\ref{eqom})
the  result can  only be a  function  of the invariants $\Pi_\perp^2$,
$p_0^2$ and $p_z^2$, and the $\gamma$-structure has
to be proportional to $\gamma\Pi_\perp$, $\gamma_0p_0$ and $\gamma_zp_z$.
We shall therefore compute the matrix elements
of Eq.~(\ref{eqom}) between the vacuum eigenstates $\Phi_\kappa$.
To be specific we use the gauge $A_\mu=(0,0,-Bx,0)$.
After computing the matrix elements of Eq.~(\ref{eqom}) we find indeed
that they are diagonal in $\kappa$ for $u_0$ and $u_z$, and
have a mixing with the first subdiagonals for $u_x$ and $u_y$.
We define $\langle u_{0,z,\pm}\rangle $ by
\begin{eqnarray}\label{u}
     \langle\Phi_{\kappa'}|\left<\frac{u_{0,z}}{u\cdot\Pi}\right>
	|\Phi_\kappa\rangle 
     &=& (2\pi)^3\delta_{\kappa',\kappa}\langle 
	u_{0,z}\rangle _\kappa~~,\\[2mm]
     \langle\Phi_{\kappa'}|\left<\frac{u_x\pm iu_y}
	{u\cdot\Pi}\right>|\Phi_\kappa\rangle 
     &=& (2\pi)^3\delta_{\kappa',\kappa\mp1}\langle u_\pm\rangle _\kappa~~,
\end{eqnarray}
and $\kappa\mp1=\{p_0,n\mp1,p_y,p_z\}$, where $n$ labels the Landau levels. 

In  the  equilibrium real-time finite   temperature formalism the free
propagator is a $2\times2$ matrix,  but  since  we shall compute  a
one-point  function  to find the chiral charge we only need  the 11-part:
\begin{equation}\label{Tprop}
	iS_F^{\beta}(p)=iS_F^0(p)
	-f_F(p_0)\left(iS_F^0(p)-iS_F^{0*}(p)\right)~~,
\ee
where $f_F(p_0)$ is the thermal distribution function. 
The  fermionic   part of  the   HTL
effective action is simply related to the inverse of the propagator by
\begin{equation}\label{LeqS}
	{\cal L}_{\rm HTL}^{\rm f}=\overline{\Psi}(x) S^{-1}(x,y) \Psi(y)~~.
\ee
The Feynman propagator ($\Pi_0=p_0+i\epsilon p_0$) is the given by
\begin{equation}\label{S0HTL}
	iS_F^0(x,y)=\langle\bfT[\Psi(x)\overline{\Psi}(y)]\rangle =
	\left<x\left|\frac{i}{\Pi\slask-m-{\cal M}_e^2
	\gamma_\mu
	\left<\frac{u^\mu}{u\cdot\Pi}\right>}\right|y\right>~~.
\ee
With the explicit expression of the inverse propagator in the Landau level
basis it is straightforward to calculate the
spectral function $\cal A$. In the lowest Landau level, which is the only
one we need to obtain the anomaly, we find for right-handed particles
\begin{eqnarray}\label{ALLL}
    \cA^R_{\rm LLL}(E,p_z)&=&
    \tr\left[{1 \over 2}(1+\g_5)\g_0\cA(E,p_z,n=0)\right]
    \nn
    &=&{1 \over2\pi i}\Bigl(S^R_{\rm LLL}(E-i\eps,p_z)
    -S^R_{\rm LLL}(E+i\eps,p_z)\Bigr)~~,\nn
    S^R_{\rm LLL}(E,p_z)&=&
    {1 \over{p_0-p_z-{\cal M}_e^2(\langle u_0\rangle -\langle u_z\rangle )}}~~.
\end{eqnarray}
\section{The chiral anomaly}\label{s:chiral}
The classical
action for massless fermions is invariant under chiral transformations,
but the corresponding chiral current is not conserved on the
quantum level due to the chiral anomaly.\cite{AdlerBJ69}
The divergence of the chiral current in 3+1 dimensions is given by
\begin{equation}\label{chcons}
    \partial_\mu j^\mu_5=
    \partial_\mu\overline{\Psi}\g^\mu\g_5\Psi=
    \frac{e^2}{16\pi^2}\varepsilon_{\mu\nu\rho\sigma}
    F^{\mu\nu}F^{\rho\sigma}~~.
\ee
Finite temperature effects
do not break chirality and, as a classical action, the HTL
effective action is still chirally invariant. 

To verify Eq.~(\ref{chcons}) explicitly 
we shall compute the produced chiral charge
$\langle Q_5\rangle=\int d^3x\langle\overline{\Psi}(x)
\g_0\g_5\Psi(x)\rangle$ in a
constant magnetic field when a parallel electric field is applied.
The chiral charge has to be defined using a gauge-invariant point splitting
regularization in the spatial $z$-direction \cite{AmbjornGP83}
\begin{eqnarray}\label{Q5}
    \langle Q^\gamma_5\rangle &=&
    \int dx\,dy\,dz\,dz'\frac{\exp[-\frac{(z-z')^2}{2\g}]}{\sqrt{2\pi\g}}
	\nn &&
    \times\langle\overline{\Psi}(x,y,z,t) \g_0\g_5\Psi(x,y,z',t)\rangle 
    \exp\left[ie\int_{z'}^{z} A_z(z'',t)dz''\right]~.
\eea
The  field
expectation values can be  related to the time-ordered Feynman Green's
function via
\begin{equation}\label{expS}
	\langle\overline{\Psi}(x)\g_0\g_5\Psi(y)\rangle=
    -i\,\tr\left[\at{S_F(y,x)}{x_0>y_0}\g_0\g_5\right]~~.
\ee
In last section we derived an explicit expression for the propagator in a
background magnetic field, and it turns out to be rather easy to include a
parallel electric field with arbitrary time dependence in the gauge
$A_\mu=(0,0,0,A_3(t))$ for the electric field. It only amounts to a phase
shift of the eigenstates (for more details see Ref.~[1]). 

Just as in vacuum the higher Landau levels ($n\geq 1$) do not contribute to
the anomaly. It can also be shown \cite{Smilga92,Elmfors97} that the purely
thermal part of the propagator (the second part of the right-hand side of
Eq.(\ref{Tprop})) does not give any contribution to the
anomaly. We are left with possible contributions from the lowest Landau level.
The relevant expectation value is
($\kappa_0=\{n=0,p_y,p_z\}$):
\begin{equation}\label{SkappaLLL}
    \tr S_F(\kappa_0)\g_0\g_5=
    \int_{-\infty}^\infty dE ~\frac{\cA^R_{\rm LLL}(E,p_z)-
      \cA^L_{\rm LLL}(E,p_z)}{p_0-E+i\eps p_0}~~,
\ee
with $\cA^{L}_{\rm LLL}(E,p_z)=\cA^{R}_{\rm LLL}(E,-p_z)$.
The produced chiral charge reduces to
\begin{equation}\label{Q5Z}
    \langle Q_5\rangle^\g _{\rm LLL}=-\frac{VeB}{4\pi^2}\int dp_z
    e^{-\frac{\g}{2}(p_z-eA_z)^2}
    \left[Z(p_z)-Z(-p_z)\right]~~,
\ee
where $Z(p_z)=\int_0^\infty dE \cA^R_{\rm LLL}(E,p_z)$.
It is the spectral weight for the right-handed positive energy solution
in the lowest
Landau level. For very large $|p_z|$ there are no collective
excitations, such as holes, but only the standard particle solution.
Therefore, $Z(p_z\goto-\infty)=0$ and $Z(p_z\goto\infty)=1$
(see Fig.~\ref{f:n0x2}), which implies
\be\label{D5fin}
    \langle Q_5(t)\rangle_{\rm LLL}=-\frac{VeB}{2\pi^2}eA_z
    =\int d^3x\int^tdt'\frac{e^2}{16\pi^2}
    \varepsilon^{\mu\nu\rho\sigma}F_{\mu\nu}F_{\rho\sigma}~~,
\ee
in agreement with Eq.~(\ref{chcons}).
\section*{Acknowledgments}
This work has been supported by the Swedish Natural Science Counsil
grant 10542-303.
It is based upon Ref.~[1] where more details can be
found. 

\end{document}